\documentclass{aastex}
\usepackage{spr-astr-addons}
\usepackage{url}

\RequirePackage{color}

\begin{document}

\title{Dynamical friction in an isentropic gas}
%\slugcomment{Not to appear in Nonlearned J., 45.}
%% Running heads
\shorttitle{Dynamical friction in an isentropic gas}
\shortauthors{Khajenabi and Dib}

\author{Fazeleh Khajenabi\altaffilmark{1} and Sami Dib\altaffilmark{2}}
\affil{Department of Physics, Faculty of Sciences, Golestan University, Basij Square, Gorgan, Iran\\f.khajenabi@gu.ac.ir}
\affil{Astrophysics Group, Blackett Laboratory, Imperial College London, London SW7 2AZ, United Kingdom\\s.dib@imperial.ac.uk}

\begin{abstract}
When a gravitating object moves across a given mass  distribution, it creates an overdense wake behind it. Here, we performed an analytical study of the structure of the flow far from object when the flow is isentropic and  the object moves  subsonically within it. We show that the dynamical friction force is the main drag force on  the object and by using a perturbation theory, we obtain the density, velocity and pressure of the perturbed flow far from the mass. We derive the expression of the dynamical friction force in an isentropic flow   and show its dependence on the Mach number of the flow and on the adiabatic index. We find that the dynamical friction force becomes lower as the adiabatic index increases. We show analytically that the wakes are less dense in our isentropic case in comparison to the isothermal ones.
\end{abstract}


%\section*{}
%\label{sec:intro}

\section{Introduction}

The motion of a gravitating mass accross a given distribution of matter creates an overdense wake behind it. In turn, this overdense region interacts gravitationally with the moving object causing it to lose momentum and kinetic energy. The ambient medium gains the momentum and the kinetic energy lost by the moving mass. The drag force exerted on the moving object is also known as the {\it dynamical friction force}. \citet{chand} studied the effects of the dynamical friction force in the collisionless systems analytically and his results has been applied to many astrophysical systems. The effects of the dynamical friction force have been investigated in the context of the evolution of massive young stellar clusters near to the Galactic center   \citep*[e.g.,][]{gerhard, kim}, the dynamical evolution of stellar clusters \citep*[e.g.,][]{portegies02}, and the migration of planets in a protoplanetary disc made of collisionless planetesimals  \citep*[e.g.,][]{poplo,muto}. N-body simulations have also shown that the dynamical friction force based on the Chandrasekhar's formula can be extended to the many-body interacting  systems \citep{inoue}.

Dynamical friction effects are also present when a gravitating mass moves accross a collisional system, i.e. a gaseous medium. The work done by the dynamical friction force may act as a heating mechanism of the gas. \citet{elzant} and \citet{fal} have  shown that the intergalactic gas in the central regions of the galaxy clusters can be heated by the motions of the galaxies. Other authors have investigated the role of the gas drag force in the context of planet migration in gaseous protoplanetary discs \citep*[e.g.,][]{ida96,hagigipour03}, the dynamics of the protostars and the stars in the stellar clusters prior to the gas removal \citep*[e.g.,][]{escala03,saiyadpour97}, and the motion of black holes on circular orbits in a gas rich galactic nuclei \citep*{kimh09,kimwt10}. The way gas drag affects the collisional behaviour of planetesimals in a gaseous disc has also been investigated by several authors \citep*[e.g.,][]{kobayashi10,ormel10}.

From a more fundamental point of view, a number of authors studied the nature of the drag force when the perturber moves supersonically in the gaseous medium \citep{dok,rud,repha}. In the steady-state subsonic case, \citet{repha} concluded that the drag force is absent. \citet{ostrik} studied the time-dependent  linear theory of the dynamical friction force and found the value of dynamical friction is not zero in contrary to the previous studies. Simulations performed by \citet{sanchez}  confirmed the analytical results of \citet{ostrik}, although other simulations find a different result (e.g., Ruffert 1996).

More recently, \citet{lee} (hereafter LS) considered the role of the dynamical friction in the subsonic case. They argued that because time does not appear explicitly in Ostriker's solution, a steady-state analysis can be applied in the subsonic case. In fact, the dynamical friction force should reach a steady state if the object travels for a sufficiently long time through a very extended medium. In all simulations of this problem, a steady-state value seems to be approached. It is true that, in linear perturbation theory, there is no dynamical friction force in the subsonic regime \citep*{ostrik}. There is also no mass accretion onto the object in linear theory, as was shown in LS. Of course, neither the force nor the accretion rate is truly zero. Both are finite if one extends the perturbation analysis to second order.  Also, LS assumed the object's radius is much smaller than the accretion radius as it is truly valid for real astronomical objects and under this assumption the relative density enhancement in the wake is not small and so they could extend their analysis to the nonlinear case.

Ostriker (1999) attempted to {\it circumvent} this zero-drag problem by assuming that the gravitational force from the moving body suddenly switches on at a certain time. There is no physical basis for this assumption. Moreover, Ostriker calculated the force by integrating over linear density perturbations in the gas. But, under her assumption that the physical size of the object is much smaller than the accretion radius, the density perturbations are nonlinear close to the object. LS approach uses perturbation theory where it is indeed valid, far from the object. We also accounted for that portion of the force arising from direct accretion of momentum, a contribution neglected by Ostriker. On the other hand, not all the numerical simulations have confirmed Ostriker's result, in particular those which assumed a small accretor obtained
a very different force (e.g., Ruffert 1996).   But LS  results showed that the friction force is given by $\dot{M} V$ (where $\dot{M}$ is the mass accretion rate onto the object/perturber and $V$  is its velocity) as long as the perturber is moving subsonically.

In the present work, we extend the approach of LS to the isentropic case. We derive the steady-state analytical solutions for the dynamical friction force when the ambient medium is isentropic. Our results show that the dynamical friction force remains equal to $\dot{M} V$ similar to the isothermal case but in our case $\dot{M}$ is the mass accretion rate when the gas is isentropic. We focus on the region far from the object and we also assume that the object's radius  is much smaller than the accretion radius. The basic equations are derived in \S~2. The perturbed equations and the boundary conditions are introduced in \S~3 as well as the mass accretion rate and its relation to the stream function are obtained in this section. In \S~4, we solve the dynamical equations for the first-order perturbed equations. In \S~5, we solve the second-order equations and the dynamical friction force is calculated in the isentropic regime. Our results are summarized in \S~6.

\section{basic equations}
The basic equations of our problem in the steady state can be written as
\begin{equation}
\nabla \cdot (\rho {\bf U}) =0,
\label{eq1}
\end{equation}
\begin{equation}
\rho ({\bf U} \cdot \nabla) {\bf U} = -\nabla P - \rho \frac{GM}{R^2},
\label{eq2}
\end{equation}
\begin{equation}
({\bf U}.\nabla)E-\frac{P}{\rho^2}({\bf U}.\nabla)\rho=0,
\label{eq3}
\end{equation}
where $\rho$, ${\bf U}$, $P$, $E$ are the density, the velocity, the pressure, and energy, respectively. The energy  $E$ is defined as $ E= 1/ (\gamma-1)(P/ \rho)$, where $\gamma$ is the adiabatic index. Here, $R$ is the radial distance in the spherical coordinates $(R,\theta,\varphi)$ whose origin is anchored on the gravitating mass $M$.  We also assume that the gas is isentropic with $\gamma>1$.  We assume that very far from the mass which travels in a straight line in the medium with velocity $V$, that density is uniform, i.e. $\lim_{R\rightarrow \infty} \rho \equiv \rho_0$. It is more convenient to introduce a set of dimensionless variables which are defined as $r=R/R_{\rm s}$, ${\bf u} = {\bf U} / c_{\rm s}$, $\varrho = \rho/\rho_{0}$ and ${ p}={ P}/P_{0}$, where $c_{\rm s}$ is the sound speed and $R_{s}$ is  the sonic radius which is defined as $R_{\rm s} \equiv GM/c_{\rm s}^2$. Equation (\ref{eq3}) is the energy equation and we also introduce the dimensionless variable  $e=E/E_{0}=p/\varrho$.  Using these new variables, equations (\ref{eq1})-(\ref{eq3}) can be rewritten as
\begin{equation}
\nabla \cdot (\varrho {\bf u})=0,
\label{eq4}
\end{equation}
\begin{equation}
\varrho ({\bf u} \cdot \nabla ) {\bf u} = - \nabla  p - \frac{\varrho}{r^2},
\label{eq5}
\end{equation}
\begin{equation}
({\bf u}.\nabla)\frac{e}{\gamma-1}-\frac{p}{\varrho^2}({\bf u}.\nabla)\varrho=0.
\label{eq6}
\end{equation}

It is clear that, in equations (\ref{eq4})-(\ref{eq6}), the radial derivatives are calculated with respect to the non-dimensional radial distance $r$. We also define $\beta = V/c_{\rm s}$, where $V$ is the velocity of the flow at the large distance from the object. Thus, properties of the flow at the distances far from the gravitating  mass $M$ are ${\bf u} = \beta {\rm e}_{\rm z}$, $\varrho = 1$, $p=1$, when $r\gg 1$. Note that in our calculations it is assumed that the system is axisymmetric ($\partial / \partial\varphi =0$). It is advantageous to introduce the non-dimensional stream function $\psi (r,\theta)$ which is defined as being
\begin{equation}
\varrho {\bf u} = \nabla \times (\frac{\psi}{ r \sin\theta} {\bf e}_{\varphi}),
\label{eq7}
\end{equation}
then the non-dimensional continuity equation (\ref{eq4}) is automatically satisfied. Thus, the components of the velocity can be written in terms of the stream function as
\begin{equation}\label{eq8}
u_{\rm r} = \frac{1}{\varrho r^2 \sin\theta} \frac{\partial\psi}{\partial\theta},
\end{equation}
and
\begin{equation}\label{eq9}
u_{\theta} = \frac{-1}{\varrho r \sin\theta} \frac{\partial\psi}{\partial r}.
\end{equation}

Having the above equations for the velocity in terms of stream function $\psi$, we are left only with the momentum equation (\ref{eq5}) and the energy equation (\ref{eq6}). Thus, we have
\begin{equation}
\varrho \left ( u_{\rm r} \frac{\partial u_{\rm r}}{\partial r} + \frac{u_{\theta}}{r} \frac{\partial u_{\rm r}}{\partial\theta} - \frac{u_{\theta}^2}{r} \right ) = - \frac{\partial p}{\partial r} - \frac{\varrho}{r^2},
\label{eq10}
\end{equation}
\begin{equation}
\varrho \left ( u_{\rm r} \frac{\partial u_{\theta}}{\partial r} + \frac{u_{\theta}}{r} \frac{\partial u_{\theta}}{\partial\theta} + \frac{u_{\rm r} u_{\theta}}{r} \right ) = -\frac{1}{r} \frac{\partial p}{\partial\theta},
\label{eq11}
\end{equation}
\begin{equation}
 \frac{1}{\gamma-1}\left(u_{\rm r} \frac{\partial e}{\partial r}+ \frac{u_{\theta}}{r} \frac{\partial e}{\partial\theta}\right)- \frac{p}{\varrho^2}\left(  u_{\rm r} \frac{\partial \varrho}{\partial r}+\frac{u_{\theta}}{r} \frac{\partial \varrho}{\partial\theta} \right )=0.
 \label{eq12}
\end{equation}
The above equations constitute our main dimensionless equations. In the next section, we perturb them using perturbation expansions.

\section{perturbation expansions}
The stream function and the density are expanded as in LS, i.e.
\begin{equation}
\psi (r,\theta) = f_{2}(\theta) r^2 + f_1 (\theta) r + f_0 (\theta) + f_{-1}(\theta) r^{-1} +  \cdots ,
\label{eq13}
\end{equation}
\begin{equation}
\varrho (r,\theta) = 1 + g_{-1} (\theta) r^{-1} + g_{-2} (\theta) r^{-2} + g_{-3} (\theta) r^{-3} + \cdots,
\label{eq14}
\end{equation}
and in our isentropic case, we need to expand the pressure as well

\begin{equation}
p(r,\theta)= 1 + h_{-1} (\theta) r^{-1} + h_{-2} (\theta) r^{-2} + h_{-3} (\theta) r^{-3} + \cdots.
\label{eq15}
\end{equation}
Considering the fact that velocity is constant far from the mass, $f_{2}(\theta)$ can be obtained as $f_{2}(\theta)=(1/2)\beta\sin^2\theta$. Since we set $\psi(r,\pi)=0$, then $f_{i}(\pi)=0$, for $i=1,0,-1,-2$, ect. The behaviour of the velocity gives us more boundary conditions as $f'_{i}(\pi)=f'_{i}(0)=0$ for $i=1,0,-1,-2$, ect., and $f_{i}(0)=0$ for $i=1,-1,-2$, ect. (for further details see LS). Substituting the above series expansions into  the main equations, we obtain a set of ordinary differential equations.

We can now derive a relation between the accretion rate and the coefficient of the density expansion. From the continuity equation (\ref{eq1}), the mass accretion rate is written as
\begin{equation}
\dot M= -2\pi\int_{0}^\pi \rho u_{R} R^2 \sin \theta d\theta.
\label{eq16}
\end{equation}
Substituting the expansion for the stream function into equation  (\ref{eq8}) for the radial velocity $u_{\rm r}$, the non-dimensional form of $\dot M$ in units of $2\pi\rho_{0}c_{s}R_{s}^2$ becomes (see LS for details)
\begin{equation}
\dot M= f_{0}(0).
\label{eq17}
\end{equation}
From equation (\ref{eq17}), we can see that the mass accretion rate depends just on the coefficient $f_{0}$ at $\theta =0$ which is a term that is independent of the radius $r$ in expansion (\ref{eq13}).

\section{First-order equations}
We can now substitute the expansions (\ref{eq13}), (\ref{eq14}), and (\ref{eq15}) into the  equations (\ref{eq10}) and (\ref{eq11}) and the energy equation (\ref{eq12}). Then, we match the coefficients of
each power of $r$. We find that the highest power of $r$ is $r^{-1}$, but all the corresponding coefficients on both sides of the equations vanish identically. So, we consider the next power of $r$ (i.e., $r^{-2}$) and the first-order equations are obtained. From the radial component of the momentum equation, we obtain
\begin{equation}
-\beta f_{1}'' -\beta f_{1} + \beta^{2} \sin\theta \cos\theta g_{-1}' + \beta^{2}\cos^2\theta g_{-1} -h_{-1}+1 =0,
\label{eq18}
\end{equation}
and the $\theta-$component of the momentum equation gives
\begin{equation}
\beta^2 \sin^{2}\theta  g_{-1}' + \beta^{2} \sin\theta \cos\theta  g_{-1}- h_{-1}'=0,
\label{eq19}
\end{equation}
and finally, the energy equation (\ref{eq12}) yields
\begin{equation}
\gamma  \sin^2\theta  g_{-1}' - \sin^2\theta h_{-1}' + \sin\theta\cos\theta(\gamma g_{-1}-h_{-1})=0.
\label{eq20}
\end{equation}

We solve the above differential equations analytically for $f_{1}$, $g_{-1}$ and $h_{-1}$. To the first-order, the solutions determine  properties of the flow around the moving object. We can integrate  equation (\ref{eq20}) simply by re-arranging its terms as
\begin{equation}
 \sin\theta \cos\theta(\gamma g_{-1} -  h_{-1} ) + \sin^2\theta \frac{d}{d\theta} (\gamma g_{-1} - h_{-1}) =0.
 \label{eq21}
\end{equation}
Introducing $W=\gamma g_{-1} -  h_{-1}$, this equation becomes $dW/d\theta = - W \cot\theta$ and its solution is $W=W_0 /\sin\theta $ where $W_0$ is an arbitrary constant to be determined from the boundary conditions. Thus,
\begin{equation}
\gamma g_{-1} -  h_{-1}=W_0 /\sin\theta
\label{eq22}
\end{equation}

From this equation, one can simply calculate  $h'_{-1}$ in terms of $g'_{-1}$  which enables us to re-write  equation (\ref{eq20}) as follows
\begin{equation}
(\beta^2 \sin^{2}\theta -\gamma) g_{-1}' + \beta^{2} \sin\theta \cos\theta  g_{-1}- W_0 \cos\theta /\sin^2\theta=0.
\label{eq23}
\end{equation}
This equation is integrable and gives us
\begin{equation}
g_{-1}(\theta)=\frac{W_{0}}{\gamma \sin\theta}+\frac{C}{\sqrt{\gamma-\beta^2 \sin^2\theta}},
\label{eq24}
\end{equation}
where $C$ is a constant of  integration. The above equation is valid for the any value of $\theta$ including $\theta = \pi$, but  equation (\ref{eq24}) goes to infinity when we set $\theta = \pi$ unless $W_{0}=0$. Substituting expression of $g_{-1}$ into equation (\ref{eq22}), we obtain  $h_{-1}$ as
\begin{equation}
h_{-1}(\theta)=\frac{C\gamma}{\sqrt{\gamma-\beta^2\sin^2\theta}}.
\label{eq25}
\end{equation}

Considering equation (\ref{eq18}) and the above equations for $g_{-1}$ and $h_{-1}$, a differential  equation for $f_{1}$ is obtained
\begin{equation}
f_{1}''+f_{1}=\frac{1}{\beta}-\frac{\gamma C}{\beta}\frac{(\gamma-\beta^2)}{(\gamma-\beta^2\sin^2\theta)^ \frac{3}{2}}.
\label{eq26}
\end{equation}
The general solution of the above equation is written as
\begin{equation}
f_{1}=\frac{1}{\beta}-\frac{C}{\beta}\sqrt{\gamma-\beta^2\sin^2\theta}+A\cos\theta+B\sin\theta.
\label{eq27}
\end{equation}
The boundary conditions help us to determine the constants $A$, $B$ and $C$. Since we have $f_{1}(\pi)=0$ and $f_{1}(0)=0$, then
\begin{equation}
1-A\beta=C\sqrt{\gamma},
\label{eq28}
\end{equation}
and
\begin{equation}
1+A\beta=C\sqrt{\gamma}.
\label{eq29}
\end{equation}
\begin{figure}
\includegraphics[scale=0.45]{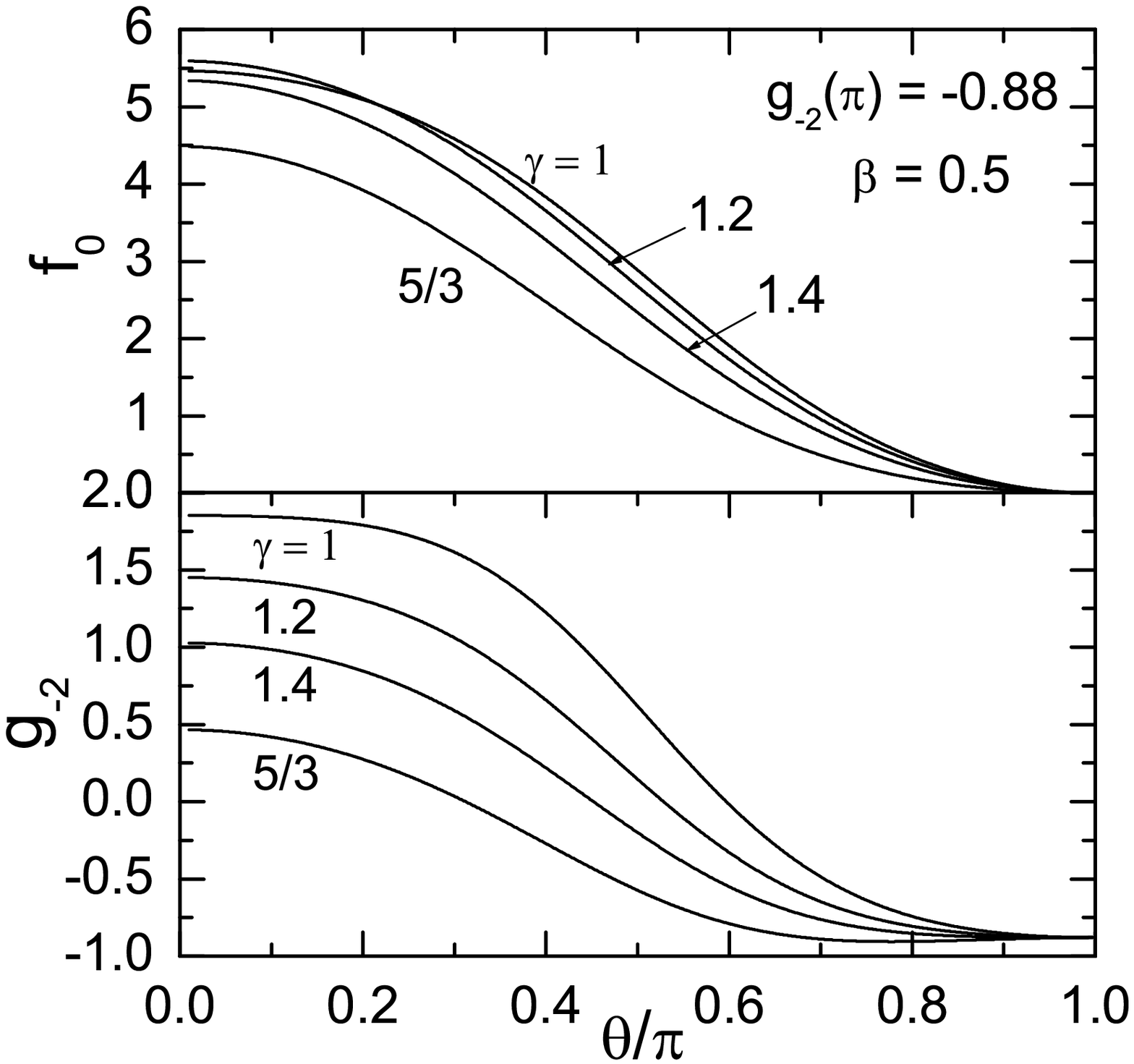}
\caption{Profiles of $f_{0}(\theta)$ and $g_{-2}(\theta )$ for different values of $\gamma$ with $g_{-2}(\pi)=-0.88$.}
\label{fig1}
\end{figure}
Therefore, the constants become $C=1/\sqrt{\gamma}$ and $A=0$. Similarly, the boundary conditions  $f_{1}'(\pi)=0$ and $f_{1}'(0)=0$ give $B=0$. So, the coefficient $f_{1}(\theta)$ can be written as
\begin{equation}
f_{1}(\theta)=\frac{1}{\beta}-\frac{1}{\beta\sqrt{\gamma}}\sqrt{\gamma-\beta^2\sin^2\theta},
\label{eq30}
\end{equation}
and also we have
\begin{equation}
g_{-1}(\theta ) = \frac{1/\sqrt{\gamma}}{\sqrt{\gamma - \beta^2 \sin^2\theta }},
\end{equation}
\begin{equation}
h_{-1}(\theta ) = \frac{\sqrt{\gamma }}{\sqrt{\gamma - \beta^2 \sin^2\theta }}.
\end{equation}
Up to now, we have derived the first-order density, pressure, and stream function perturbations analytically for an isentropic flow. For the isothermal case (i.e., $\gamma=1$), our first-order solutions reduce to the first-order solutions of LS.

\begin{figure}
\includegraphics[scale=0.45]{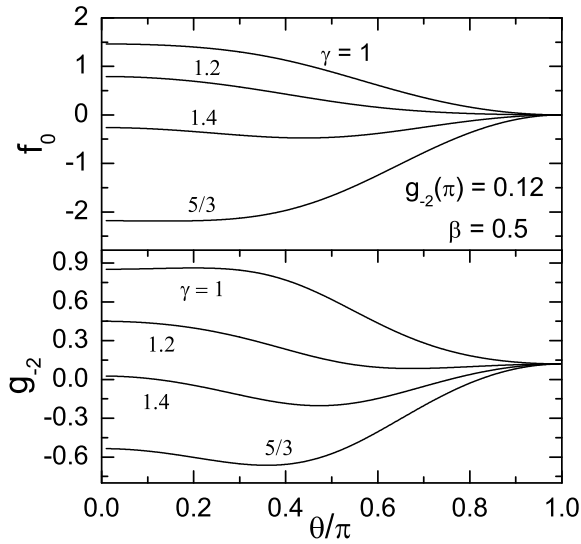}
\caption{Profiles of $f_{0}(\theta)$ and $g_{-2}(\theta )$ for different values of $\gamma$ with $g_{-2}(\pi)=0.12$.}
\label{fig2}
\end{figure}

\begin{figure}
\includegraphics[scale=0.45]{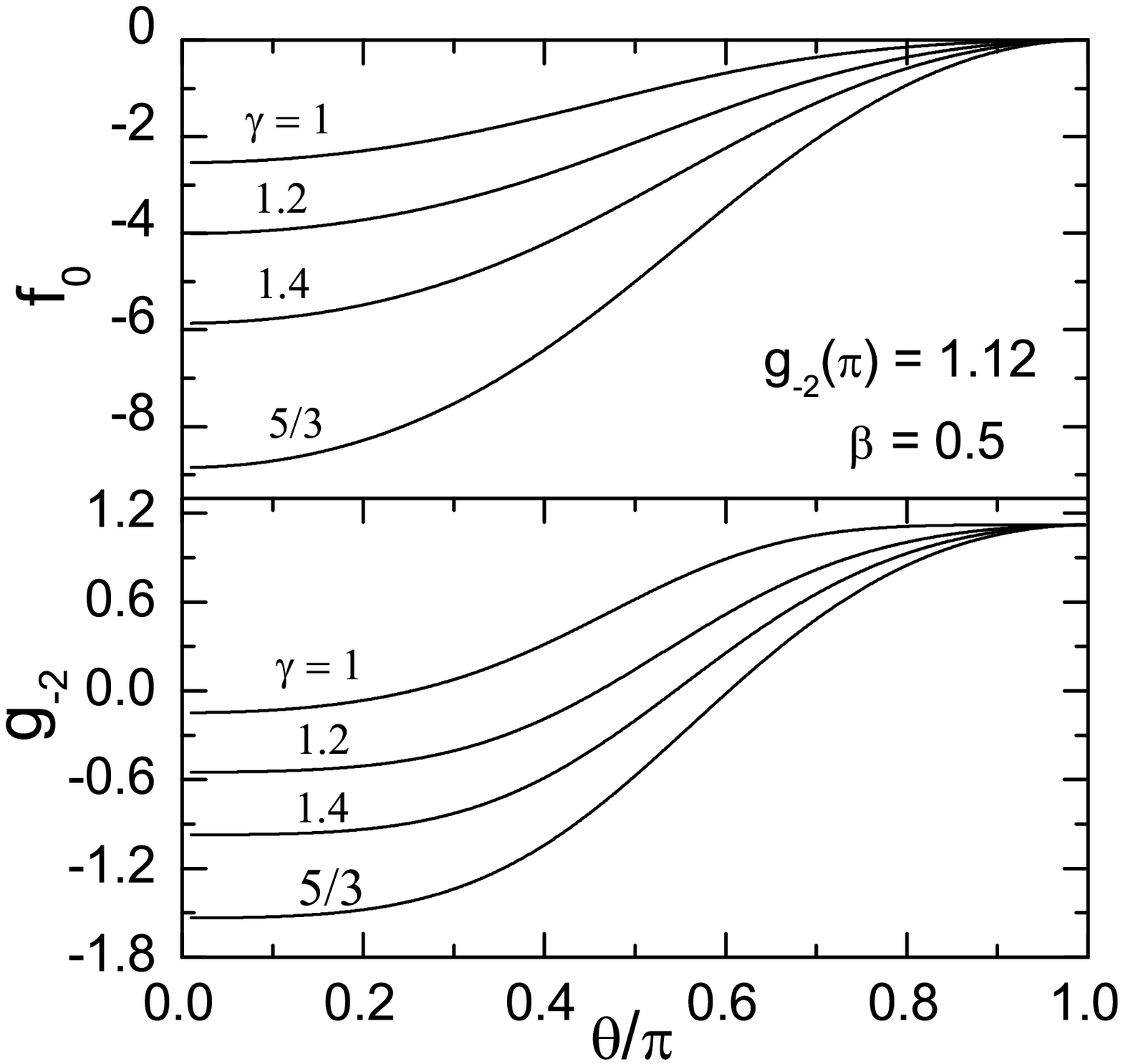}
\caption{Profiles of $f_{0}(\theta)$ and $g_{-2}(\theta )$ for different values of $\gamma$ with $g_{-2}(\pi)=1.12$.}
\label{fig3}
\end{figure}

\begin{figure}
\includegraphics[scale=0.45]{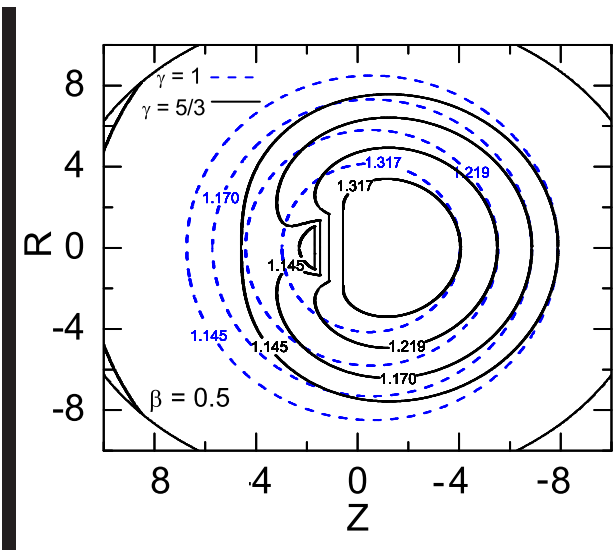}
\caption{The density contours correspond to  $\gamma=5/3$ (solid) and $\gamma=1$ (dashed) for the cases  with $\beta=0.5$ and $g_{-2}(\pi)=1.12$.}
\label{fig4}
\end{figure}

\section{second-order equations}
Now, we equate the coefficients of the next power of $r$ in the components of the momentum equation and the energy equation in order to obtain  the second-order equations. From the radial component of the momentum equation, we obtain
\begin{displaymath}
-\beta f_{0}''-\beta \cot\theta f_{0}'+\beta^2\cos\theta\sin\theta g_{-2}'+
\end{displaymath}
\begin{equation}
2\beta^2\cos^2\theta g_{-2}-2h_{-2} = {\cal A}_{1} + {\cal A}_{2} + {\cal A}_{3} + {\cal A}_{4},
\label{eq31}
\end{equation}
where all the terms on the righthand side depend on the first order variables as
\begin{equation}
{\cal A}_{1}= \frac{f_{1}^2}{\sin^2 \theta} - \frac{f_{1}f_{1}' \cos\theta }{\sin^{3}\theta} + \frac{(f_{1}')^2}{\sin^2 \theta} + \frac{f_{1} f_{1}''}{\sin^2 \theta},
\label{eq32}
\end{equation}
\begin{displaymath}
{\cal A}_{2} = \beta f_1 g_{-1} - 2\beta f_{1}' g_{-1} \cot\theta - \beta f_{1} g_{-1}' \cot\theta -
\end{displaymath}
\begin{equation}
\beta f_{1}' g_{-1}' + \beta f_{1}'' g_{-1},
\label{33}
\end{equation}
\begin{equation}
{\cal A}_{3} = - 3 g_{-1},
\label{eq34}
\end{equation}
\begin{equation}
{\cal A}_{4} = 2g_{-1}h_{-1}.
\label{eq35}
\end{equation}
Also, there is another equation from the $\theta-$ component of equation of motion,
\begin{displaymath}
-\beta f_{0}'-\beta^2\sin^2\theta g_{-2}'+h_{-2}' -2\beta^2\cos \theta \sin \theta g_{-2}
\end{displaymath}
\begin{equation}
 = {\cal B}_{1}+{\cal B}_{2}+{\cal B}_{3},
 \label{eq36}
\end{equation}
where
\begin{equation}
{\cal B}_{1}= f_1^{2} \frac{\cot\theta}{\sin^2\theta} - \frac{f_{1} f_{1}'}{\sin^2\theta},
\label{eq37}
\end{equation}
\begin{equation}
{\cal B}_{2} = \beta f_{1} g_{-1} \cot\theta + \beta f_{1}' g_{-1} + 2\beta f_{1} g_{-1}',
\label{eq38}
\end{equation}
\begin{equation}
{\cal B}_{3}=-2g_{-1} h_{-1}'.
\label{eq39}
\end{equation}
Finally the last equation is obtained from the energy equation,
\begin{displaymath}
\beta \sin^2\theta h_{-2}'-\gamma \beta \sin^2\theta g_{-2}'+2\beta \sin\theta \cos\theta h_{-2}-2\gamma\beta\sin\theta
\end{displaymath}
\begin{displaymath}
\times\cos\theta g_{-2}=(\gamma-1)\beta \sin\theta\cos\theta g_{-1}h_{-1}+\gamma \beta \sin^2\theta g_{-1}'h_{-1}
\end{displaymath}
\begin{equation}
-\beta\sin^2\theta g_{-1}h_{-1}'-f_{1}h_{-1}'-h_{-1}f_{1}'+\gamma  f_{1}g_{-1}'+\gamma f_{1}'g_{-1}.
\label{eq40}
\end{equation}

The right-hand side of the equations (\ref{eq31}), (\ref{eq36}) and (\ref{eq40}) are determined analytically because we have already found the coefficients  $f_1$, $g_{-1}$ and $h_{-1}$ analytically.

One can rewrite the left-hand side of equation (\ref{eq40}) as,
\begin{displaymath}
\beta \sin^2\theta \frac{d}{d\theta}(h_{-2}-\gamma g_{-2})+2\beta \sin\theta\cos\theta(h_{-2}-\gamma g_{-2})=
\end{displaymath}
\begin{equation}
\frac{d}{d\theta} [\beta \sin^2\theta(h_{-2}-\gamma g_{-2})].
\label{eq41}
\end{equation}
After substituting the expressions of $g_{-1}$, $h_{-1}$, and $f_{1}$ into the right-hand side of  equation (\ref{eq40}) and considering equation (\ref{eq41}), we can integrate equation (\ref{eq40}) as follows
\begin{equation}
\beta^2 \sin^2\theta (h_{-2}-\gamma g_{-2})=\frac{1}{2}\frac{\gamma(\gamma-1)}{\beta(\gamma-\beta^2 \sin^2\theta)}+K,
\label{eq42}
\end{equation}
where $K$ is a constant of  integration. This equation is valid over the entire range of $\theta$ including $\theta=0$. Substituting $\theta=0$ into equation (\ref{eq42}), one can obtain

\begin{equation}
\frac{1}{2}\frac{\gamma(\gamma-1)}{\beta\gamma}+K=0,
\label{eq43}
\end{equation}
and so
\begin{equation}
K=-\frac{\gamma-1}{2\beta}.
\label{eq44}
\end{equation}
Now, equation (\ref{eq42}) becomes
\begin{equation}
h_{-2}-\gamma g_{-2}=\frac{\gamma-1}{2\beta}\frac{1}{(\gamma-\beta^2 \sin^2\theta)}.
\label{eq45}
\end{equation}

Substituting the analytical expressions of $f_{1}$ and $g_{-1}$ into the righthand side of equations (\ref{eq31}) and (\ref{eq36}) and using equation (\ref{eq45}), the following coupled differential equations for $f_{0}$ and $g_{-2}$ are obtained
\begin{displaymath}
-\beta f_{0}''-\beta \cot\theta f_{0}'+\beta^2\cos\theta\sin\theta g_{-2}'+(2\beta^2\cos^2\theta-2\gamma) g_{-2}
\end{displaymath}
\begin{equation}
= \frac{\gamma-1}{\beta}\frac{1}{D}+\frac{1}{\sqrt\gamma}(\frac{\sqrt\gamma}{D}-\frac{3}{\sqrt D}+\frac{2}{\sqrt \gamma+\sqrt D}),
\label{eq46}
\end{equation}
and
\begin{displaymath}
-\beta f_{0}'+(\gamma-\beta^2\sin^2\theta )g_{-2}'-2\beta^2\cos \theta \sin \theta g_{-2}=\beta^2\sin \theta\cos\theta
\end{displaymath}
\begin{equation}
 (\frac{1-\gamma}{\beta D^2}-\frac{2}{D^2}+\frac{2}{\sqrt \gamma  D^\frac{3}{2}}-\frac{1}{\gamma D}+\frac{1}{\gamma(\sqrt \gamma+\sqrt D)^2}),
 \label{eq47}
\end{equation}
where $D=\gamma-\beta^2\sin^2\theta $.

Before solving the second-order equations numerically, we obtain a relation between the mass accretion rate and the second-order coefficient of the density expansion. The left-hand side of the second-order equation (\ref{eq47}) is re-written as
\begin{displaymath}
-\beta f_{0}'+(\gamma-\beta^2\sin^2\theta )g_{-2}'-2\beta^2\cos \theta \sin \theta g_{-2}
\end{displaymath}
\begin{equation}
=\frac{d}{d \theta}(-\beta f_{0}+D g_{-2}).
\label{eq48}
\end{equation}
Having the above equation for the left-hand side of equation (\ref{eq48}) and the fact that $D$ is
an even function of $\theta$, we can integrate both side of  the equation (\ref{eq47}) from $\theta=\pi$ to $0$ which yields,
\begin{equation}
\left(-\beta f_{0}+D g_{-2}\right)_{\theta=\pi}=\left(-\beta f_{0}+D g_{-2}\right)_{\theta=0},
\label{eq49}
\end{equation}
where the right hand side of the equation vanishes and since $f_{0}(\pi)=0$ and $D(\pi)=D(0)=\gamma$, we obtain
\begin{equation}
f_{0}(0)=\frac{\gamma}{\beta}(g_{-2}(0)-g_{-2}(\pi))
\label{eq50}
\end{equation}
If we set $\gamma=1$, the above equation  reduces  to  equation (46) in LS. Equation (\ref{eq50}) indicates a relation between the mass accretion rate and the second-order coefficient of density perturbation. The accretion rate is proportional to the difference of the second-order density perturbation in the upstream and downstream flows and the proportionality constant depends on the velocity of the object as well as on the value of $\gamma$. Although equation (\ref{eq50}) gives us a physical insight of the accretion in terms of the coefficient $g_{-2}$, $\gamma$ and $\beta$, we can not determine the accretion rate only based on this equation. Because we don't have a physical condition on the coefficient of $g_{-2}$ at $\theta =0$. Instead, we prescribe the accretion rate (i.e., $f_0 (0)$) from an accretion model.

One should solve the two main equations  (\ref{eq46}) and  (\ref{eq47}) numerically subject to the mentioned boundary conditions. We note that  for the isothermal flow (i.e., $\gamma=1$), our second-order equations (\ref{eq46}) and (\ref{eq47}) reduce to the equations (35) and (36) of LS. In order to solve the equations as a two point boundary value problem, the equations are integrated numerically from $\theta=\pi$ to $\theta=0$. Also, we need three initial values to start numerical integration from $\theta =0$, two of which were introduced earlier, i.e. $f_{0}(\pi)=f'_{0}(\pi)=0$. We have  $g_{-2}(\pi)$ as the third boundary condition, but there is no constraint on its value. In order to obtain the physical value of $g_{-2}(\pi)$, we follow the approach of LS, in which the mass accretion rate is prescribed  via a physical model to give us $f_{0}(0)=\dot{M}$. Then, we try to find $g_{-2}(\pi)$ so that after solving the equations numerically, the prescribed value for $f_0$ at $\theta =0$ is obtained.  The standard Bondi-Holye accretion model was used by LS in their model for determining the accretion rate. But we can not use the Bondi-Hoyle accretion rate, because our flow is isentropic.  \citet{fog} did an extensive analysis to  calculate  appropriate relations for the accretion rate when the gas is isentropic.  Their non-dimensional mass accretion rate is
\begin{equation}
\dot M (\beta)=2(\frac{1}{2})^{\frac{\gamma+1}{2(\gamma-1)}}(\frac{4}{5-3\gamma})^{\frac{5-3\gamma}{2(\gamma-1)}}
(1+(\frac{\gamma-1}{2})\beta^{2})^{\frac{5-3\gamma}{2(\gamma-1)}}.
\label{eq51}
\end{equation}
\begin{figure}
\includegraphics[scale=0.45]{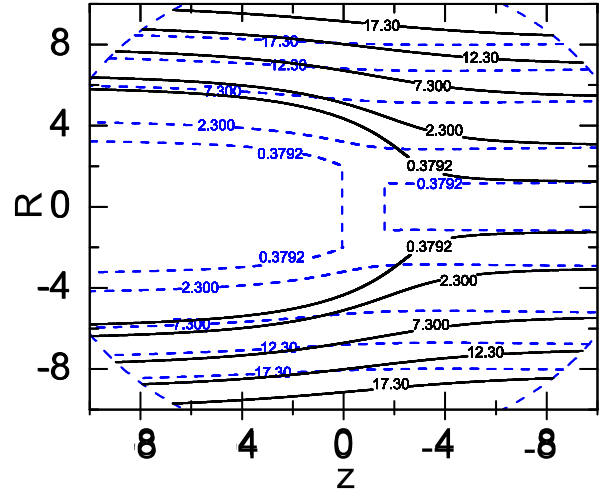}
\caption{The streamlines  correspond to $\gamma=5/3$ (solid) and $\gamma=1$ (dashed)  for the cases with $\beta=0.5$ and $g_{-2}(\pi)=1.12$.}
\label{fig5}
\end{figure}
\begin{figure}
\includegraphics[scale=0.45]{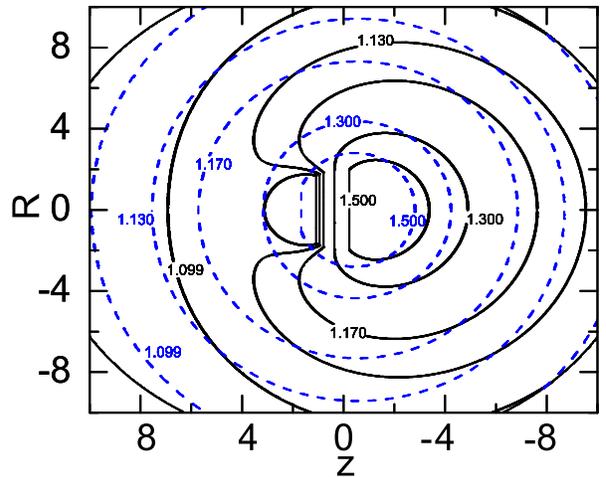}
\caption{The pressure contours correspond to $\gamma=5/3$ (solid) and $\gamma=1$ (dashed) for the cases  with $\beta=0.5$ and $g_{-2}(\pi)=1.12$.}
\label{fig6}
\end{figure}
Using  equations (\ref{eq51}) and (\ref{eq17}), we obtain  value of  $f_{0}(0)$ for given  $\beta$ and $\gamma$ and it helps us to the calculate  the proper value of $g_{-2}(\pi)$.

\begin{figure}
\includegraphics[scale=0.45]{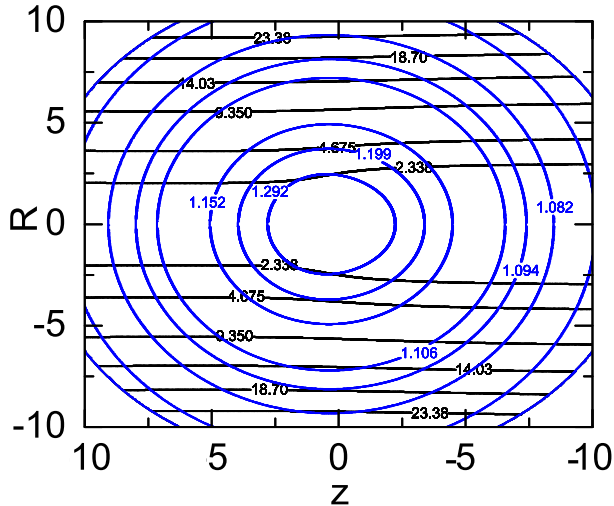}
\caption{The streamlines and the density contours for a realistic case based on equation (\ref{eq51}) for the accretion rate corresponding to $\gamma = 1.4$ and $\beta = 0.5$. Here, we have $f_{0}(0)=\dot{M}=1.25$ and $g_{-2}(\pi ) = -0.15$.}
\label{fignew}
\end{figure}

The behaviour of the second-order coefficients $f_{0}(\theta)$ and $g_{-2}(\theta)$ are shown in Figures ~\ref{fig1}, \ref{fig2}, and \ref{fig3}. For $\beta=0.5$, we adopted  three fiducial boundary values of $g_{-2}(\pi)$ (i.e., $-0.88,0.12,1.12$) which allows us to make an easier comparison of our results for an isentropic flow to  the solutions of LS in the isothermal case. Each curve is labeled by the adiabatic index $\gamma$. Profiles of the second-order coefficients are modified  significantly when $\gamma$ varies from 1 to the larger values. Both $f_{0}(\theta)$ and $g_{-2}(\theta)$ decrease for all values of $g_{-2}(\pi)$. The density and the pressure contours and the streamlines  are also shown in Figures  \ref{fig4}, \ref{fig5} and \ref{fig6} corresponding to the two cases where $\gamma=1$ and $\gamma=5/3$. For a more physical case, we used equation (\ref{eq51}) to obtain the accretion rate corresponding to the input parameters $\gamma = 1.4$ and $\beta =0.5$. Then, we have $f_{0}(0)=\dot{M}=1.25$ and numerically it is found $g_{-2}(\pi ) = -0.15$. Figure \ref{fignew} shows the streamlines and the density contours for this case.

Now, we are in a situation where we can derive the dynamical friction force for the isentropic problem. In order to do so, we calculate the rate of momentum in the $z-$direction that is transferred from the the object to the ambient medium. Considering  a sphere of radius $R$ centred on the object, the amount of the total momentum transferred through the surface of this sphere is independent of its size as long as it lies outside the wake (LS). Thus, the total dynamical friction force  consists of the kinetic and the static parts as follows:
  \begin{equation}
  F_{\rm kin}=-2\pi \int_{0}^{\pi } \rho u_{\rm R} u_{\rm z} R^2 \sin\theta d\theta,
  \label{eq52}
  \end{equation}
and,
  \begin{equation}
  F_{\rm static}=-2\pi \int_{0}^{\pi } P R^2 \cos\theta \sin\theta d\theta.
  \label{eq53}
  \end{equation}
Then, the non-dimensional  dynamical friction force $F$ becomes
\begin{equation}
F= -\int_{0}^{\pi } \varrho u_{\rm r} u_{\rm z} r^{2} \sin\theta d\theta - \int_{0}^{\pi } p \cos\theta \sin\theta d\theta,
\label{eq54}
\end{equation}
where the unit of force is equal to $2\pi \rho_0 c_{\rm s}^2 R_{\rm s}^2$.

Now, we can substitute the corresponding expansions into equation (\ref{eq54}). We find that all terms with positive powers of $r$ vanish upon integration, but the terms which are independent of $r$ survive. Therefore,
\begin{equation}\label{eq:poly-force}
F = - \int_{0}^{\pi } \left [ \sin\theta \cos\theta (h_{-2} - \beta^2 g_{-2}) + \beta (1+\cos^{2}\theta ) f_{0}' \right ] d\theta.
%\label{eq55}
\end{equation}
Obviously, we have $h_{-2}(\theta )= g_{-2}(\theta )$ for an isothermal flow and so, the above relation for the friction force reduces to equation (57) of LS, i.e.,
\begin{equation}\label{eq:iso-force}
F=-\int_{0}^{\pi} \left((1-\beta^2)\sin\theta\cos\theta g_{-2}+\beta(1+\cos^{2}\theta)f'_{0}\right) d\theta.
%\label{eq56}
\end{equation}
\begin{figure}
\includegraphics[scale=0.45]{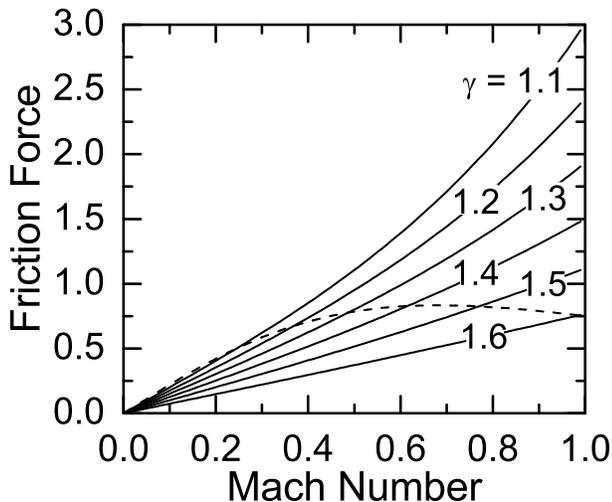}
\caption{The dimensionless friction force $F$ as a function of Mach number. Each curves is labeled by its value of $\gamma$. The dashed curve  corresponds to the LS solution.}
\label{fig7}
\end{figure}
LS showed analytically that equation (\ref{eq56}) is integrable. We also follow the same approach for evaluating the integral relation (\ref{eq55}) for our isentropic flow. Like LS, the flow is irrotational and so the $\phi -$ component of the vorticity vanishes which leads to
\begin{equation}
\beta f_{0}'' - \beta \cot\theta f_{0}' - \beta^2 \sin\theta \cos\theta g_{-2}' + 2 \beta^2 \sin^{2}\theta g_{-2} = \frac{1-\sqrt{ D}}{ D}.
\label{eq57}
\end{equation}
If we add the above equation to the equation (\ref{eq46}) and multiplying by $-(1/2) \sin\theta \cos\theta $, we obtain
\begin{equation}
\sin\theta \cos\theta (h_{-2} - \beta^2 g_{-2}) + \beta \cos^2 \theta f_{0}' = - \sin\theta \cos\theta { H} ({ D}),
\label{eq58}
\end{equation}
where ${ H}({ D})$ is a function of ${ D}$. Now, we can split equation (55) as
\begin{displaymath}
F= -\int_{0}^{\pi} \beta f_{0}'d\theta
\end{displaymath}
\begin{equation}
- \int_{0}^{\pi}\left[\sin\theta\cos\theta(h_{-2}-\beta^2 g_{-2})+\beta\cos^2\theta f_{0}'\right ]d\theta.
\label{eq59}
\end{equation}
Having equation (\ref{eq57}), the second term of the above equation vanishes and we obtain
\begin{equation}
F = \beta f_{0} (0) = \beta \dot{M} ,
\label{eq60}
\end{equation}
where $\dot M$ is the mass accretion rate in an isentropic flow (i.e., equation (\ref{eq51})). Exactly the same relation is valid for the case of isothermal flows as demonstrated by LS. We can now study the effect of the adiabatic index $\gamma$ on the dynamical friction force in Figure \ref{fig7}. This figure displays the dynamical friction force as a function of the Mach number for different values of the adiabatic index. The dashed line shows the solution derived by LS for isothermal flows. This figure shows that the dynamical friction force decreases when $\gamma$ increases. Dependence of the dynamical friction force  on the $\gamma$  becomes more significant when Mach number has a larger value. However, one should note that in our model our  the accretion rate monotonically increases in the
subsonic regime. But in the formula used by LS which is actually based on the Bondi-Holye model, the accretion rate monotonically decreases.  This fact accounts for the divergence of all our curves from their result at intermediate Mach number.

\section{Conclusion}

In the present investigation, we considered a point-mass object moving through an isentropic gas and the structure of the flow far from the object and the resulting dynamical friction force calculated semi-analytically. In fact, our study is a direct generalization of the LS approach for the isothermal gas to the isentropic case. Using perturbation expansions for the density,  pressure and the stream function, we obtained the first-order and the second-order differential equations describing the excited wakes when the object's velocity is  subsonic. We restricted our analysis to the steady state case.
In the subsonic regime, we proved analytically that the dynamical friction force is $\dot M V$, where $\dot M$ is mass accretion rate of the isentropic flow. Our results show that a larger  value of the adiabatic index corresponds to a  reduction of the friction force for a given velocity. Our analytical solutions for the isentropic case show that the wakes are less dense  in comparison to the isothermal ones because the isentropic flow is less compressible than the isothermal case. This fact has also been pointed out by LS, but our study is the first to address this issue semi-analytically.

\section*{Acknowledgments}

We are grateful to Steven Stahler and Aron Lee for their helpful comments and suggestions that improved the paper. F. K. acknowledges support of the Golestan University. S. D. is supported by STFC grant ST /H00307X/1, and acknowledges the additional support provided by a Santander Mobility Award and the hospitality of the Institut de Ci\`{e}ncies del Cosmos in Barcelona and the Instituto de %
Astrofis\'{i}ca de Andalucia in Granada.

\bibliographystyle{mn2e}
\bibliography{referenceKH_v2}

\end{document}